\documentclass[prc,amsmath,twocolumn,showpacs,superscriptaddress]{revtex4}
\usepackage{graphicx}
\usepackage{amssymb}
\usepackage{enumerate}
\usepackage{verbatim}

\begin{document}

\title{Adiabatic versus Faddeev for (d,p) and (p,d) reactions}

\author{F.M. Nunes}
\affiliation{National Superconducting Cyclotron Laboratory, Michigan State University, East Lansing, MI 48824, USA}
\affiliation{Department of Physics and Astronomy, Michigan State University, East Lansing, MI 48824-1321}
\author{A.~Deltuva}
\affiliation{Centro de F\'{\i}sica Nuclear da Universidade de Lisboa, P-1649-003 Lisboa, Portugal}

\date{\today}

\begin{abstract}
The finite range adiabatic wave approximation provides a practical method to analyze (d,p) or (p,d)
reactions, however until now the level of accuracy obtained in
the description of the reaction dynamics has not been determined. In this work, we perform
a systematic comparison between the finite range adiabatic wave approximation
and the exact Faddeev method.
We include studies of $^{11}$Be(p,d)$^{10}$Be(g.s.) at $E_p=$5, 10 and 35 MeV;
$^{12}$C(d,p)$^{13}$C(g.s.) at $E_d=$7, 12 and 56 MeV and $^{48}$Ca(d,p)$^{49}$Ca(g.s.) at $E_d=$19, 56 and 100 MeV.
Results show that the two methods agree within $\approx 5\%$ for a range of beam energies ($E_d \approx 20-40$ MeV) but differences increase significantly for very low energies and for the highest energies. Our tests show that ADWA agrees best with Faddeev when the angular momentum transfer is small $\Delta l=0$ and when the neutron-nucleus system is loosely bound.
\end{abstract}



\maketitle
\section{Introduction}

There is growing interest in using (d,p) reactions as a tool to extract single particle properties
of nuclei away from stability. Since the early studies on $^{11}$Be \cite{be11pd}, an increasing number
of transfer experiments have been performed to obtain detailed structure information.
This includes experiments at ISOL facilities, performed around the Coulomb barrier, such as the recent one
on $^{132}$Sn \cite{sn132dp}, but also measurements at higher energy in fragmentation facilities, such as
the recent systematics on Ar isotopes \cite{ar-exp}. Since the information extracted from these
experiments relies on reaction models, it is critical to validate these models and assess their uncertainties.

There are a number of aspects that need to be considered when validating reaction theories.
At present, reaction theories for (d,p), applicable to nuclei with mass $A > 10$,
require a reduction of the many-body problem to a few-body problem. A consequence of this first point is the introduction of effective interactions between
the composite constituents, the so called optical potentials and single particle binding potential. These are the physical inputs to the problem. In addition, there are often approximations in solving the few-body problem. It is this last aspect that is the focus of the present work.

For forty years, the tradition has been to use the Distorted Wave Born Approximation (DWBA) to
extract spectroscopic information from (d,p) or (p,d) reactions \cite{book}. Even though still in use, this method has been challenged
repeatedly. Just in the last decade, a variety of reaction theory studies have been carried out.
In \cite{liu04}, it is shown that the choice of the deuteron optical potential in DWBA introduces a very
large ambiguity which can be significantly reduced when using the zero-range ADiabatic Wave Approximation
developed by Johnson and Soper(ZR-ADWA) \cite{soper} built on the nucleon-target optical potentials.
In ZR-ADWA deuteron breakup is taken into account to all orders, while making a zero-range approximation
for the deuteron. The formalism for a finite range version of the adiabatic wave method (FR-ADWA) was developed in \cite{tandy},
which is also built on nucleon optical potentials.
Perhaps even larger than the uncertainties in optical potentials, the ambiguity
introduced by the single particle potentials which describes the many-body overlap function has also
been a focus of study \cite{muk05,pang06}. Target excitation is known to be relevant for
specific cases and should be considered case by case \cite{delaunay,pang06}.

To add to the long list of reaction theories,  one should still consider
the continuum discretized coupled channel method (CDCC) \cite{cdcc} and the Faddeev-type method for transition operators, usually
referred to as the Alt-Grassberger-Sandhas (AGS) method \cite{Alt67}. Although the CDCC method is probably most widely used
to describe breakup (e.g. \cite{nunes99}), the CDCC wavefunction has also been used in the context of (d,p) reactions
(e.g. \cite{deltuva07,moro09}). The AGS method including the Coulomb interaction,
originally developed for describing few-nucleon reactions \cite{deltuva:05c}, has recently been
extended to handle (d,p) and (p,d) nuclear reactions \cite{deltuva09}. The CDCC method takes breakup effects
into account to all orders without making any further approximations (in some ways,  ADWA is an approximate version of CDCC),
but Faddeev goes a step further in that transfer channels are also included explicitly in the expansion.
The Faddeev calculations represent thus the exact solution
to the full three body problem when rearrangement channels are present.
While both CDCC and AGS make less approximations compared to DWBA and ADWA, they are computationally demanding.
Moreover, their technical implementations are limited, for example, the present treatment
of Coulomb in the AGS method has been successfully applied only
 to nuclei with charge $Z<30$ \cite{deltuva09c}.
Finally, and most importantly, these two methods are based on a more complex expansion of the wavefunction, in such a way that the final cross section
depends not only on the many-body overlap of interest, but on non-trivial interferences between many overlap functions
including overlaps with states in the continuum. As a consequence, simpler approximate methods are often preferred by experimentalists.

Recently, a systematic study \cite{nguyen10} using the finite range version of the Adiabatic wave approximation (FR-ADWA) \cite{johnson-ria,tandy} found large finite range effects in (d,p) reactions, specially
at the higher energies. The transfer cross section in FR-ADWA depends on the overlap function of interest,
and not on a complicated superposition of terms with many overlap functions.
Also important, the method is not computationally expensive and is of practical use for the non expert.
Nevertheless, one needs to appreciate that FR-ADWA relies on a Sturmian expansion which is truncated to first order
(first order should not be confused with DWBA, since the adiabatic method relies on an entirely different expansion).
This first order truncation of the exact solution of the three-body problem
has never been systematically tested before.
FR-ADWA calculations in \cite{nunes11}  are used to analyze $^{34,36,46}$Ar(p,d) data at $33$ MeV/u and
compared to Faddeev to determine the error in the treatment of the reaction dynamics.
Discrepancies between FR-ADWA were found to vary considerably (6-19\%). The work in \cite{nunes11} calls for a better
understanding of the range of validity of FR-ADWA.

The aim of the present work is exactly to determine
the range of validity of ADWA (we drop FR from now on, since all ADWA calculations presented
here are finite range). In order to do that, a systematic comparison between ADWA and Faddeev is performed.
We cover a wide range of beam energies, light and intermediate mass
nuclei, well bound and loosely bound systems. We choose energies for which there is data
to ensure that the reaction theory performs sensibly, although our focus is on comparing
two theories starting from the same three-body Hamiltonian, and not dwell on detailed comparisons
with data. In section II, we provide some key aspects of ADWA and Faddeev, in section III
results are presented and discussed. Finally in Section IV conclusions are drawn.

\section{Theoretical description}
\label{theory}

The starting point for both ADWA and Faddeev is the three-body Hamiltonian for $n+p+A$:
\begin{equation}
{\cal H}_{3B}= T_\mathbf{r} + T_\mathbf{R}  + U_{nA} + U_{pA} + V_{np}\;.
\label{h3b}
\end{equation}
The interactions between the nucleons and the composite target $A$ ($U_{pA}$ and $U_{nA}$) should contain an
imaginary term representing the absorption or transfer to other channels not explicitly included.
We will focus the discussion on the reaction A(d,p)B to the ground state (g.s.) of $B=n+A$ (note that the formalism for (p,d) in prior form
is identical to that for (d,p) in post form presented here \cite{moro09}).

Following the work on the zero-range ADWA \cite{soper}, Johnson and Tandy \cite{tandy}
introduced a full finite range version considering the square-integrable Sturmian expansion. The Sturmian basis
is complete within the range of the interaction, and is defined through:
\begin{equation}
(T_r + \alpha_i V_{np}) S_i(\vec{r}) = -\varepsilon_d S_i(\vec{r})
\label{sturm}
\end{equation}
for each given state corresponding to the number of nodes $i=0,1,2,...$. The inner product
between Sturmian states is defined by
$\langle S_i |  V_{np} | S_j \rangle = - \delta_{ij}$,
and at large distances, all basis states decay exponentially according to the deuteron binding energy.
The three-body wavefunction is now expanded in terms of $S_i$:
\begin{equation}
\Psi^{(+)}(\vec r,\vec R) = \sum_{i=0}^{\infty} S_i(\vec{r}) \chi_i(\vec{R}).
\label{wf3b}
\end{equation}
This form for the three-body wavefunction is then introduced in the three-body Schr\"odinger equation
for scattering $H_{3B} \Psi=E\Psi$ which, when imposing the appropriate boundary conditions, allows one
to calculate $\chi_i(\vec{R})$. This procedure leads to a non-trivial coupled channel
equation \cite{tandy,nguyen10} which could be solved exactly \cite{laid}.

Simplicity is recovered when considering only the first term in the expansion Eq.(\ref{wf3b}).
\begin{equation}
\Psi_{AD}^{(+)}(\vec r,\vec R) =  S_0(\vec{r}) \chi_0^{AD}(\vec{R}).
\label{wf3bad}
\end{equation}
Then the coupled-channel equation reduces to an optical model type equation with distorting potential
\begin{equation}
U_{AD}(R)=  -{\langle S_0(\vec{r}) | V_{np} (U_{nA}+U_{pA}) | S_0(\vec{r}) \rangle}, \,
\label{jtpot2-eq}
\end{equation}
where, apart from a normalization factor, $S_0$ is the ground state wavefunction of the deuteron.

The three-body wavefunction Eq.(\ref{wf3bad}) is now inserted in the post-form T-matrix for the (d,p) process \cite{moro09}:
\begin{equation}
T = \langle \phi_{nA}\chi^{(-)}_{pB}| V_{np} + \Delta_{rem} | \Psi_{AD}^{(+)}\rangle\, ,
\label{tad-eq}
\end{equation}
where $\chi^{(-)}_{pB}$ is the outgoing proton wave, distorted by $U_{pB}^*$, $\phi_{nA}$ is the wavefunction
of the final bound state, generated by a binding potential $V_{nA}$, and the remnant term is
$\Delta_{rem}=U_{pA} - U_{pB}$.

Even if we were not truncating the Sturmian basis, one would expect the method to be at
its best when remnant contributions are small, because then the resulting cross section
is only sensitive to short distances between the neutron and proton, where the basis is complete.
However, it is not clear that this condition is sufficient when the additional simplification
Eq.(\ref{wf3bad}) is introduced.

\vspace{1cm}

As opposed to ADWA, in the Faddeev approach, elastic scattering, transfer, and breakup channels
are treated on equal footing. Therefore, it represents the exact solution to the problem once
a three-body Hamiltonian Eq.(\ref{h3b}) is defined.
In the Faddeev method, the wavefunction is expanded in an over-complete basis,
involving all three Jacobi coordinates \cite{book}. In the AGS method \cite{Alt67},
one starts from the Faddeev formalism and arrives at coupled integral equations
for the transition operators
\begin{equation}\label{eq:AGS}
 T_{\beta \alpha} = {} (1 - {\delta}_{\beta \alpha}) G_0^{-1} 
     + \sum_{\gamma=1}^{3} (1 - {\delta}_{\beta \gamma}) t_\gamma  G_0 T_{\gamma \alpha}
\end{equation}
whose on-shell matrix elements
$\langle\psi_{\beta}|T_{\beta \alpha}|\psi_{\alpha}\rangle$
are scattering amplitudes and therefore lead directly to the  observables.
In Eq.(\ref{eq:AGS})
$G_0 = (E+i0-H_0)^{-1}$ is the free resolvent, $E$ being the available
three-particle energy in the
center of mass (c.m.) system and $H_0$ the free Hamiltonian.
The  two-particle transition matrix is a solution of the Lippmann-Schwinger
equation
\begin{equation}
t_{\gamma} = v_{\gamma} + v_{\gamma} G_0 t_{\gamma}
\end{equation}
where $v_{\gamma}$ is the potential for the pair $\gamma$; we use
the odd-man-out notation.
The channel states $|\psi_{\gamma}\rangle$ for  $\gamma = 1,2,3$ are the
eigenstates of the corresponding channel Hamiltonian $H_\gamma = H_0 + v_\gamma$
with the energy eigenvalue $E$; thus, $|\psi_{\gamma}\rangle$ is a product of
the bound state wave function for pair $\gamma$ and a plane wave
with fixed on-shell momentum
corresponding to the relative motion of particle $\gamma$ and pair $\gamma$
in the initial or final state.

The Coulomb interaction is an additional complication in the AGS method;
nevertheless, when only two particles are charged and charge numbers $Z$ are not too large,
the Coulomb interaction
has been successfully included \cite{deltuva09,deltuva09c} using the method of screening and renormalization
\cite{deltuva05,alt:80a}.

We solve the AGS equations in the momentum-space partial-wave framework
\cite{deltuva03a}. The Coulomb screening radius needed for the convergence of the results
increases with $Z$ thereby increasing also the number of partial waves that have to be included.
Such calculations are not only  time consuming but
for very high angular momenta the partial-wave expansion may even become unstable.
This limits the application of the technique with respect to $Z$; so far
no calculations have been performed beyond $Z=30$  \cite{deltuva09c}.
Another limitation of the method comes from the Pade summation technique for solving
the AGS integral equation iteratively \cite{chmielewski:03a}; usually it is hard to achieve convergence
for heavier nuclei at low energies.
For more details on the numerical techniques we refer the reader to
\cite{deltuva05,deltuva03a,chmielewski:03a,deltuva05d}.


\begin{figure}[!t]
{\centering \resizebox*{0.42\textwidth}{!}{\includegraphics{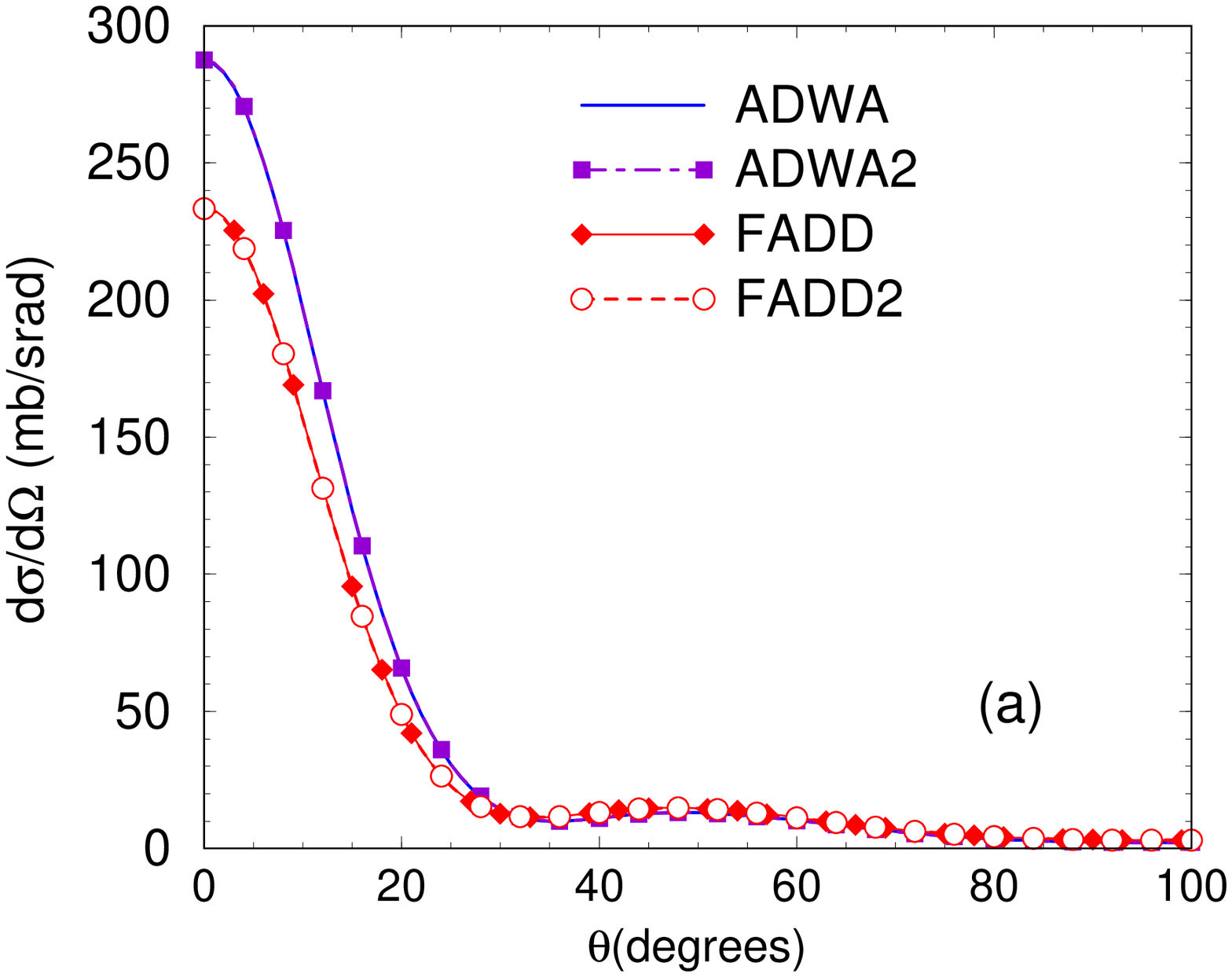}}} \\
{\centering \resizebox*{0.42\textwidth}{!}{\includegraphics{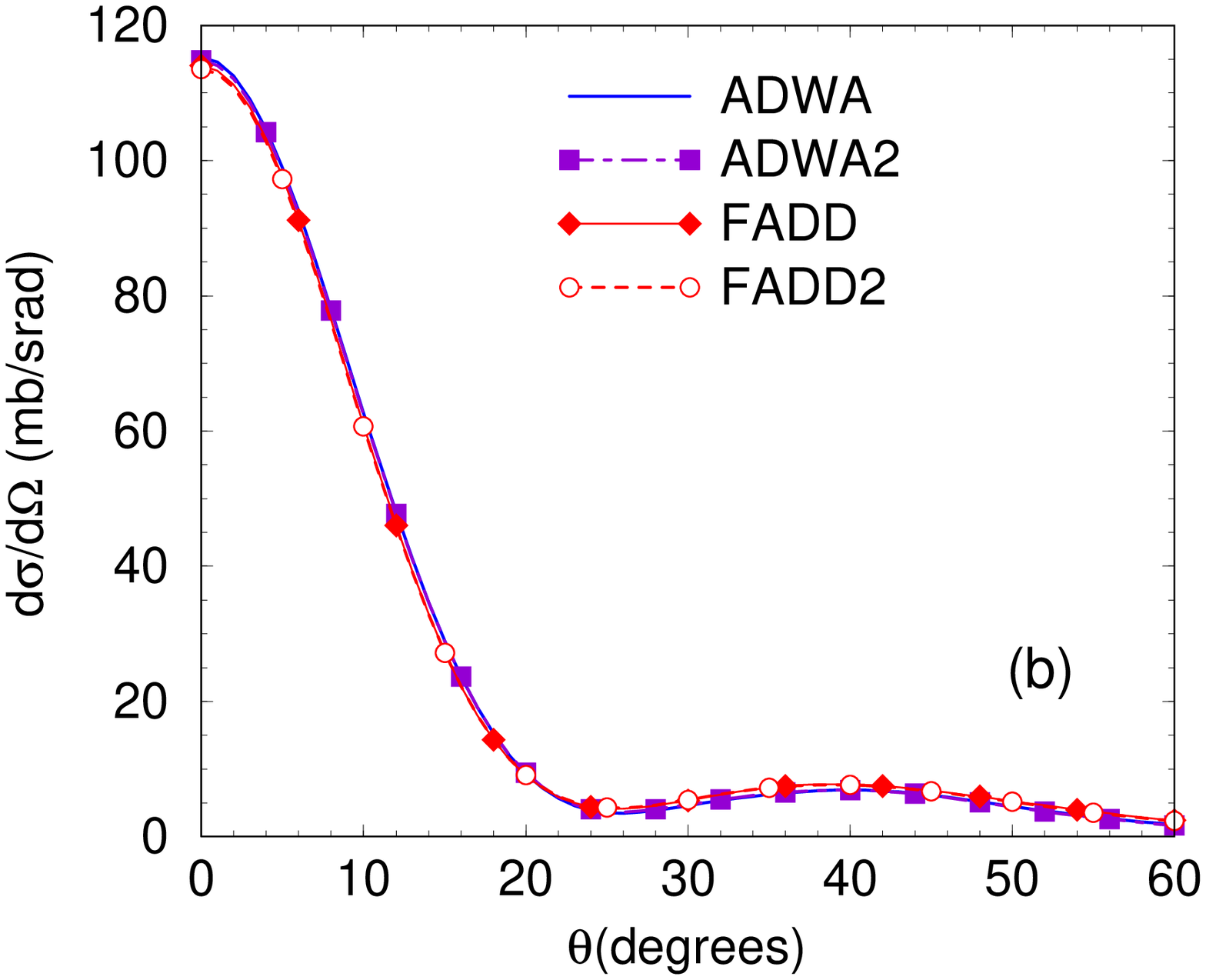}}} \\
{\centering \resizebox*{0.42\textwidth}{!}{\includegraphics{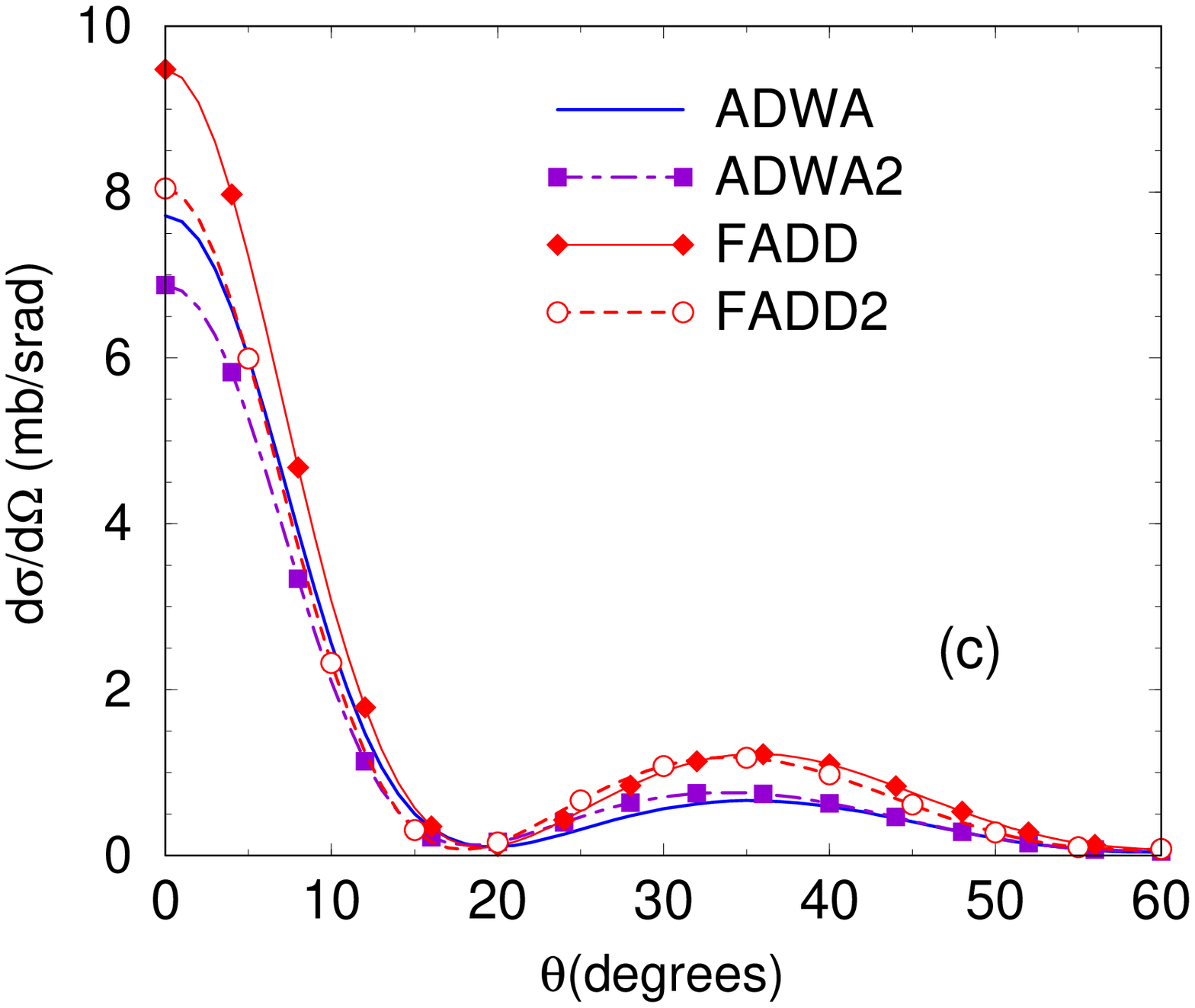}}} \\
\caption{\label{fig-be11} (Color online) Angular distributions for $^{11}$Be(p,d)$^{10}$Be:
(a) $E_p = 5$ MeV, (b) $E_p = 10$ MeV, and (c) $E_p = 35$ MeV.}
\end{figure}
\section{Results}
\label{results}

We perform Faddeev AGS and finite range ADWA calculations for a number of cases
that have been repeatedly studied: $^{11}$Be(p,d)$^{10}$Be(g.s); $^{12}$C(d,p)$^{13}$C(g.s) and $^{48}$Ca(d,p)$^{49}$Ca(g.s.).
Since we are interested in establishing the range of validity of ADWA, we explore these at several beam energies.
For nearly all cases we choose beam energies for which data is available and we make sure that the present theories are able to provide a fair description of the angular distributions.
However, since in this work we are concerned with testing ADWA against the exact Faddeev AGS theory, we do not include data in the plots.

One of the main inputs to these calculations are the pair interactions.
In both calculations, a realistic NN interaction is used. Since the ADWA is performed in coordinate space, and
a local form of the NN interaction with LS coupling is needed, the Reid interaction \cite{rsc} is used. 
The more recent momentum based CDBonn \cite{bonn} is straightforward to use in AGS calculations, contrary to potentials
in coordinate space. These two NN interactions reproduce the low-energy NN phase shifts, the deuteron binding energy,
and the quadrupole moment. In fact, apart from the short distances, the deuteron densities produced with Reid and
CDBonn are identical, this being true also for the 3N and 4N systems. We have verified that the use of different NN interactions
does not induce differences in the predicted (d,p) calculations by repeating both FADD and ADWA calculations with a simple Gaussian
interaction, that reproduces low-energy phase shifts and the deuteron binding energy. We concluded that the transfer cross sections 
are not sensitive to the details of this interaction but the consistency between the $V_{np}$ used to obtain the deuteron 
wavefunction and the interaction in the transfer operator (Eq.\ref{tad-eq}). This same conclusion was found in
previous works \cite{deltuva09}.

For the nucleon optical potentials we use the global parameterization CH89 \cite{ch89}. As all global optical potentials,
the parameters of CH89 are functions of the beam energy.
In ADWA, when determining $\Psi_{AD}$ in Eq.(\ref{tad-eq}), half the deuteron energy is used for $U_{nA}$ and $U_{pA}$ , whereas
the exit proton energy is used for the auxiliary potential $U_{pB}$ (for calculating $\chi_{pB}$ and the remnant contribution). The same $U_{pA}$ used for $\Psi_{AD}$  is used
in the transfer operator.
A real interaction $V_{nA}$ is used to calculate the
bound state $\phi_{nA}$: it  consists of a Woods-Saxon plus spin orbit form with the Woods-Saxon depth adjusted to reproduce the experimental binding energy of the nucleus under study.
In all cases we fix the geometry of the Woods-Saxon, the radius $r=1.25$ fm and diffuseness $a=0.65$ fm, as well as
the spin orbit parameters $V_{so}=6$ MeV, $r_{so}=1.25$ fm, $a_{so}=0.65$ fm.
Thus, strictly speaking, since the neutron interaction is different in the
incoming and outgoing channel, two different three-body Hamiltonians are used in ADWA.

In the Faddeev calculation, energy dependence in the interactions
can introduce orthogonality errors. It is thus important to make a wise choice for the energy at which $U_{pA}$ is to be calculated. Our standard choice is to fix $U_{pA}$ at the energy in the proton channel (FADD), however we will also
show results using half the deuteron energy (FADD2).
Concerning the neutron-target interaction, for the partial wave where a bound state exists,
the same $V_{nA}$ interaction is used as for ADWA.  For all other partial waves, the neutron interaction is $U_{nA}$, taken from CH89
at half the deuteron energy \cite{ch89}. In our standard Faddeev calculation (FADD), there are two subtle differences in the interactions
as compared to ADWA:
i) $U_{pA}$ for the deuteron channel is obtained at the proton energy, and ii) for the partial waves where a neutron bound state exists,
the neutron scattering potential $U_{nA}$ has no absorption.
Because we have seen a larger dependence on the transfer cross sections to
$U_{pA}$ than to $U_{nA}$ in previous calculations (see e.g. Fig.10 of \cite{deltuva-E}),
we believe the uncertainty due to i) dominates the overall uncertainty.

\begin{figure}[!t]
{\centering \resizebox*{0.42\textwidth}{!}{\includegraphics{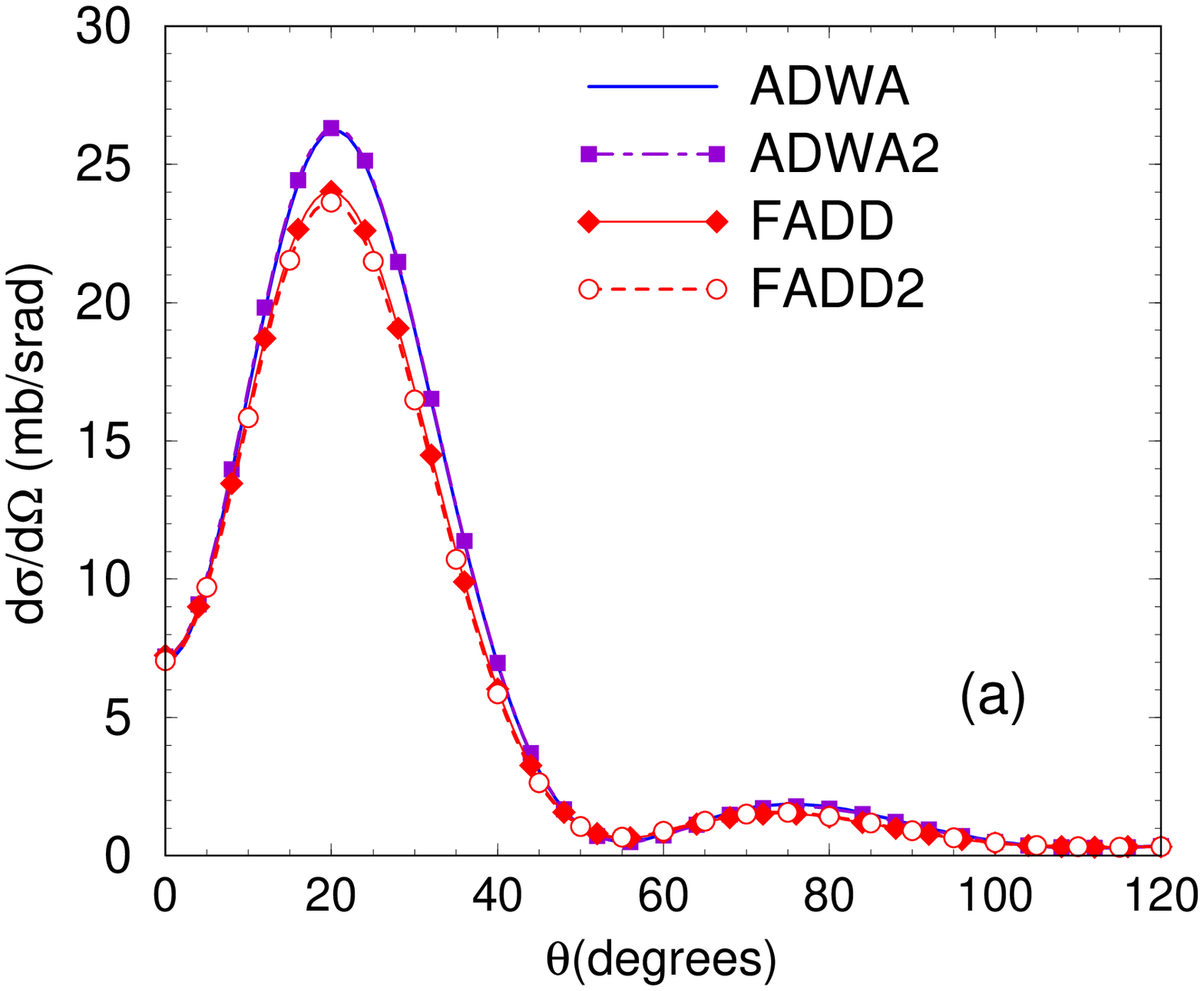}}} \\
{\centering \resizebox*{0.42\textwidth}{!}{\includegraphics{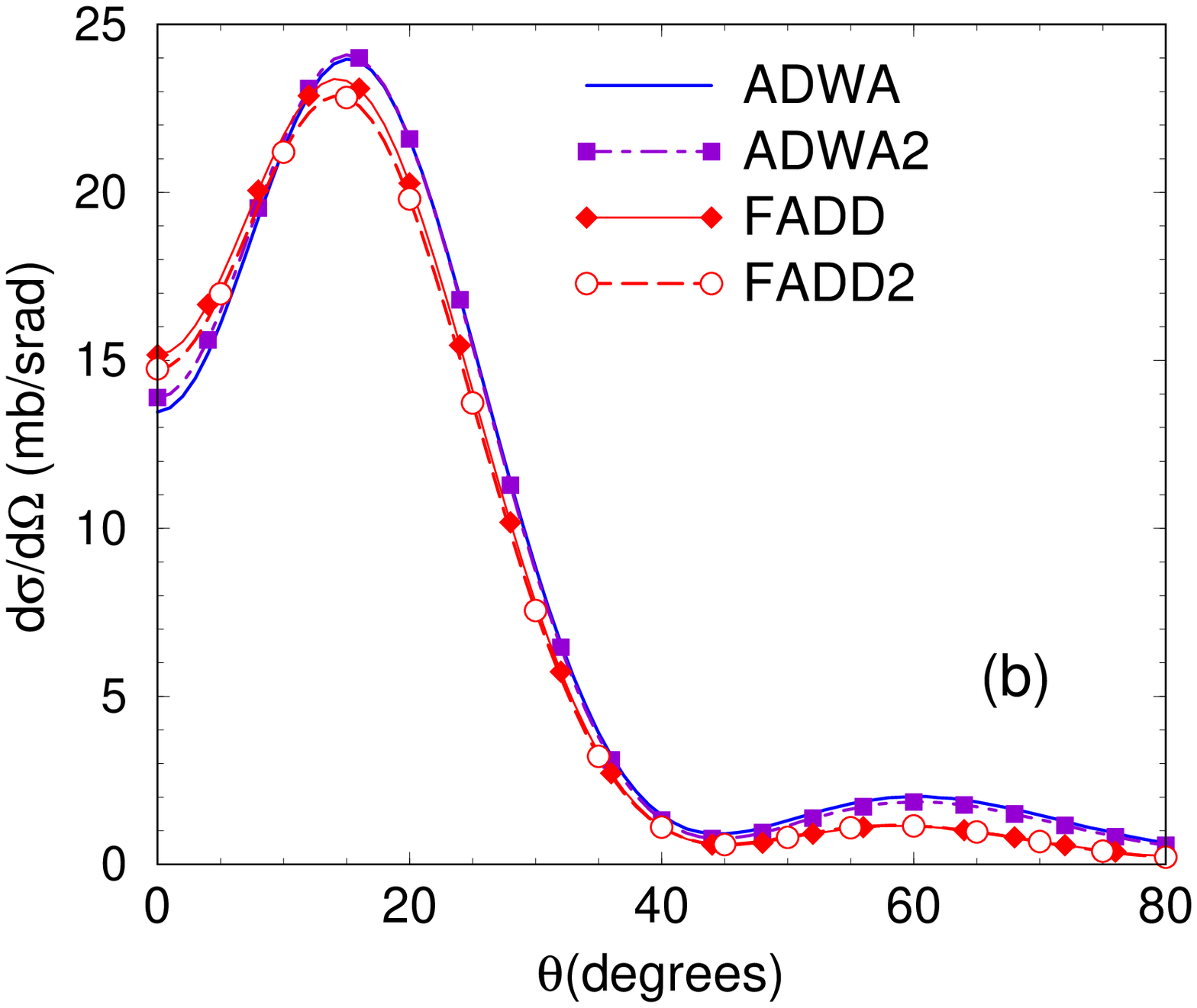}}} \\
{\centering \resizebox*{0.42\textwidth}{!}{\includegraphics{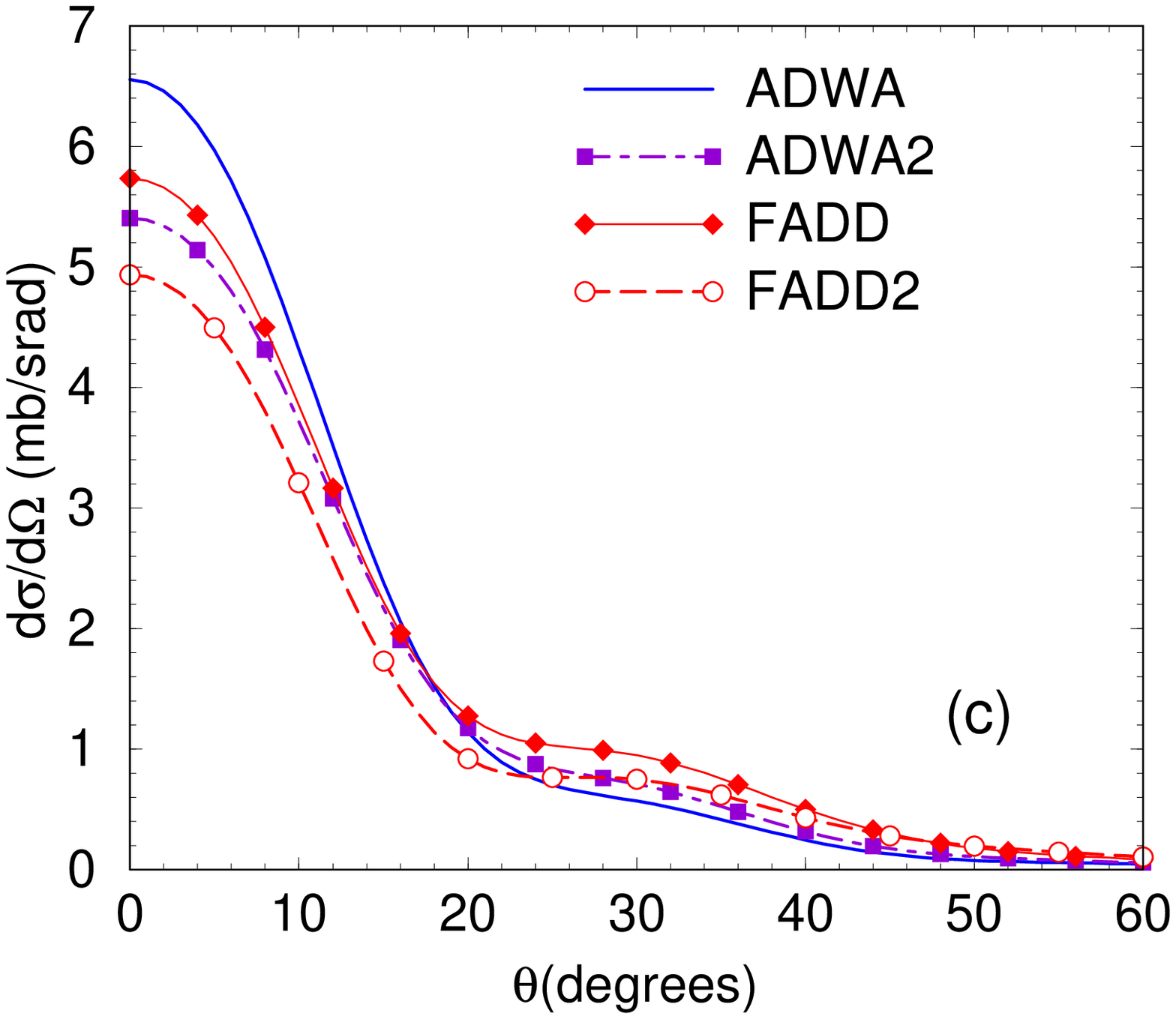}}} \\
\caption{\label{fig-c12} (Color online) Angular distributions for $^{12}$C(d,p)$^{13}$C:
(a) $E_d = 7.15$ MeV, (b) $E_d = 12$ MeV and  (c) $E_d = 56$ MeV}
\end{figure}
\begin{figure}[!t]
{\centering \resizebox*{0.42\textwidth}{!}{\includegraphics{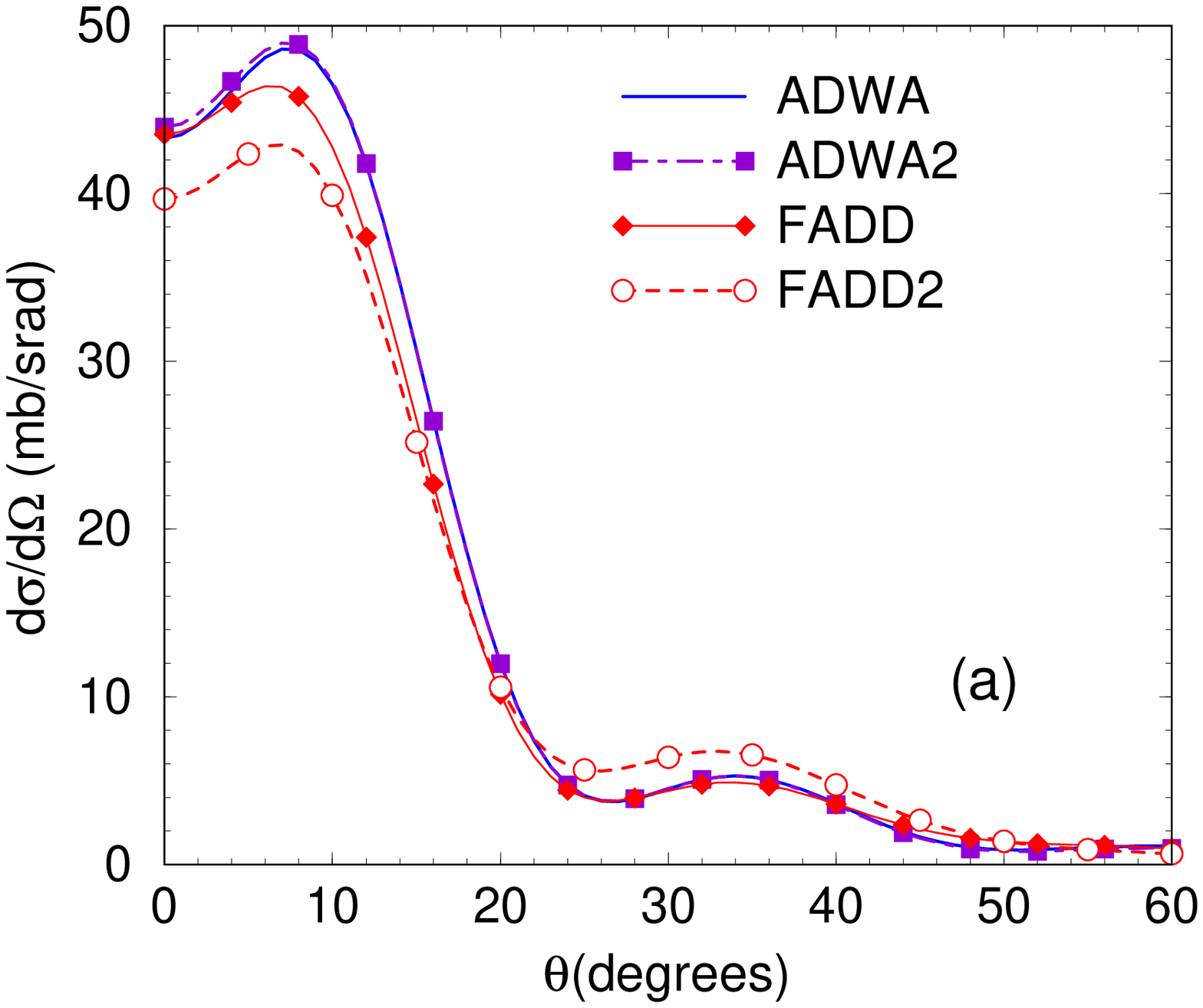}}} \\
{\centering \resizebox*{0.42\textwidth}{!}{\includegraphics{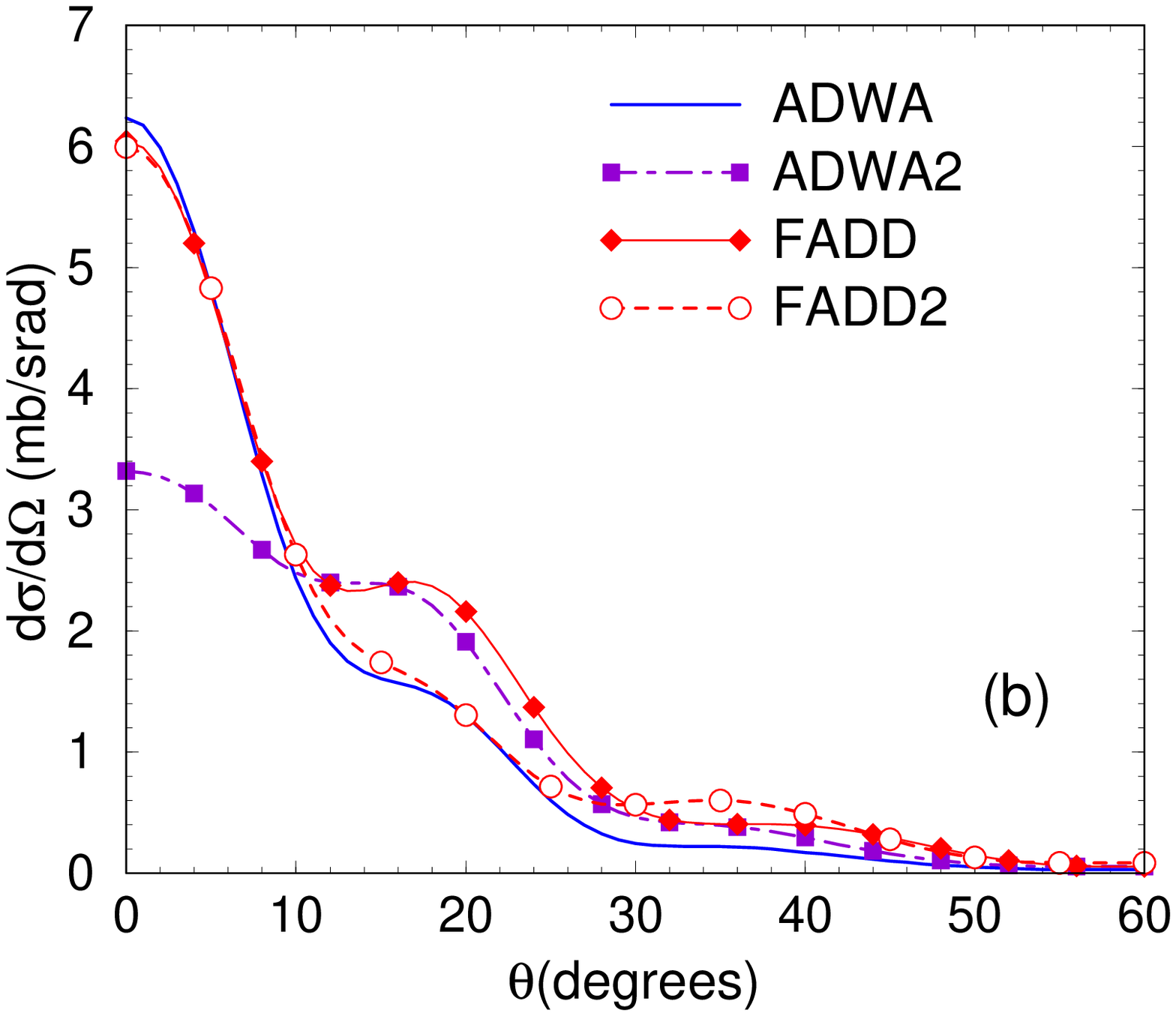}}} \\
{\centering \resizebox*{0.42\textwidth}{!}{\includegraphics{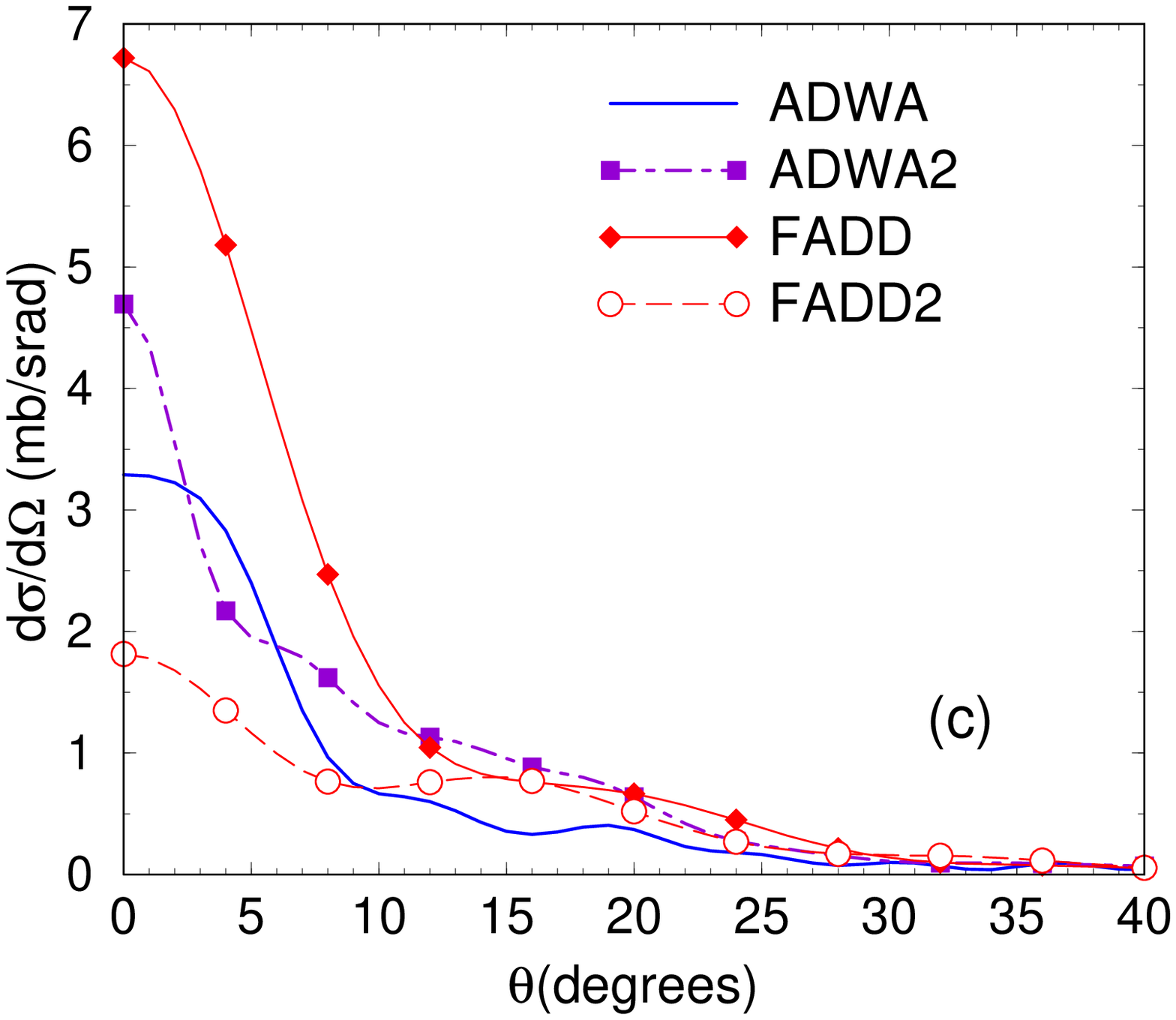}}} \\
\caption{\label{fig-ca48} (Color online) Angular distributions for $^{48}$Ca(d,p)$^{49}$Ca:
(a) $E_d = 19.3$ MeV, (b) $E_d = 56$ MeV and (c) $E_d = 100$ MeV.
}
\end{figure}

In Fig.\ref{fig-be11} we present the angular distribution for $^{11}$Be(p,d)$^{10}$Be,
at $E_p = 5$ MeV, $E_p = 10$ MeV, and $E_p = 35$ MeV. The  blue lines are the results obtained
with ADWA and the red curves are those with Faddeev. In order to understand the subtle difference in the interactions
included in the Faddeev calculations compared to ADWA, we perform two test calculations:
i) a Faddeev calculation using the proton optical potential calculated at half the deuteron energy
(red open circles labeled FADD2) and ii) an ADWA calculation where the initial proton distorted wave is calculated with an optical potential
determined at half the deuteron energy (purple filled squares labeled ADWA2). The comparison between FADD and FADD2 provides
an estimate of the sensitivity to the choice of energy at which the proton-target optical potential is evaluated.
If the two methods were exact, the comparison between FADD2 and ADWA2 would provide an estimate of the effect of replacing, in the
calculation of $\Psi_{AD}$, the appropriate neutron optical potential by a real bound state potential in the partial wave where a bound state exists. Of course ADWA2 is not
exact, and therefore the comparison of FADD2 and ADWA2 contains both the difference due to the neutron optical potential
(which should be small) but also a genuine difference between the treatment of the three-body dynamics.
Note that if $\Psi_{AD}$ were exact, the T-matrix of Eq.(\ref{tad-eq}) should provide cross sections independent of
the auxiliary potential $U_{pB}$ (used in $\chi_{pB}$ and $\Delta_{rem}$). This was recently demonstrated within CDCC \cite{moro09}.
Therefore, differences between ADWA and ADWA2 indicate inaccuracies in ADWA.

First of all, there are differences between Faddeev and ADWA cross sections at the lower and higher energy. At $E_p=10$ MeV
the agreement is perfect. Comparing FADD with FADD2, we find that only at the higher energy is there a dependence in
the choice of energy at which the proton interaction is calculated, and then Faddeev calculations using the proton interaction
at half the deuteron energy become slightly closer to the corresponding ADWA results (compare FADD2 and ADWA2). At $E_p=35$ MeV,
ADWA and ADWA2 differ by $11$\% at the peak, which indicates that the truncation to the first term in ADWA is insufficient
at these energies.

In Fig.\ref{fig-c12} the same observable is shown for $^{12}$C(d,p)$^{13}$C at $E_d = 7.15$ MeV, $E_d = 12$ MeV and
$E_d = 56$ MeV. Differences between the Faddeev and ADWA are smallest at $E_d=12$ MeV. As for $^{11}$Be, only at the highest
beam energy are the results sensitive to the choice of energy at which the proton optical potential is defined.
In this case, calculating the final proton distorted wave with $U_{pB}$ evaluated at half the
deuteron energy slightly improves the agreement (compare ADWA2 versus FADD2). This is not to say that adiabatic calculations for (d,p)
should be performed using $U_{pB}$ at half the deuteron energy. The standard ADWA makes the correct choices. Here ADWA2 only
helps to understand which part of the disagreement with Faddeev comes from the slight differences in the proton interaction.

Our last case  $^{48}$Ca(d,p)$^{49}$Ca is shown  in Fig.\ref{fig-ca48}. Angular distributions were calculated
at $E_d = 19.3$ MeV, $E_d = 56$ MeV and $E_d = 100$ MeV. The comparison between ADWA and Faddeev follows the same
trend as in the previous cases: good agreement at around 10 MeV/u, with large deterioration at very high energies.
It was not possible to obtain Faddeev solutions for energies lower that $19.3$ MeV due to lack of convergence
of the Pade summation technique. Oddly, for the reactions at $56$ MeV, while ADWA showed a very strong dependence
on the choice of $U_{pB}$, the Faddeev results only exhibit a dependence on the proton optical potential
around the second peak. At 100 MeV, the angular distribution
is extremely sensitive to the choice of the proton energy used to determine the proton optical potential parameters
for both ADWA and FADD. This introduces large ambiguity in the comparison of ADWA and Faddeev, at 100 MeV.

Overall, we find the differences between ADWA and Faddeev to be below 10 \% except for the lowest and highest energies.
Our results suggest that there is an optimum beam energy
(around 10-20 MeV/u) where ADWA is at its best, differing from Faddeev by only a few percent.
Apart from the 100 MeV $^{48}$Ca(d,p), changes introduced in the shape of the angular distributions by the adiabatic approximation
are small. A quantitative summary of our results, plus additional detail in our calculations, is given in Table \ref{errors}.
Here $\Delta_{F-AD}$ is the percentage difference between cross sections from FADD and ADWA, relative to the FADD cross section determined at the first peak of the distribution (corresponding to $\theta$). To study the effect of the choice of the energy at which the proton interaction is determined, we also show $\Delta_{F-F2}$, the percentage difference between Faddeev using the energy in the proton channel to determine $U_{pA}$ (FADD) and that using half the deuteron energy (FADD2). This represents the ambiguity in the three-body Hamiltonian for the Faddeev calculations, and can be used as an uncertainty in the comparison. It is not important for low and intermediate beam energies, and therefore in this beam-energy region, the agreement between Faddeev and ADWA is robust.
For completeness, the identical quantity for ADWA  $\Delta_{AD-AD2}$ is also shown (where the standard ADWA is compared with that
where the proton distorted wave is calculated with CH89 using half the deuteron energy ADWA2). As the exact T-matrix should
be independent on the choice of the auxiliary potential, $\Delta_{AD-AD2}$ provides an internal warning sign that
the approximation in ADWA is not adequate. Consistent with the comparison with Faddeev, the effect becomes significant
for the higher beam energies.

In Section II, we pointed out that the Sturmian expansion is expected to be good in the region of $V_{np}$ but
not for large neutron-proton distances. For this reason, we do not expect ADWA to work when the contribution
of the remnant term in Eq.(\ref{tad-eq}) is large. The last column of Table II contains the percentage difference
between the original ADWA and the ADWA calculation neglecting the remnant term, relative to the original ADWA.
The only case for which the remnant contribution is large is for the high energy reaction with $^{11}$Be
and then, indeed, ADWA performs poorly in comparison with Faddeev. However in the other cases, it is not
the remnant contribution that is responsible for the disagreement ADWA versus Faddeev. Since the largest disagreement comes at the highest
beam energies, this suggests that the main source for the ADWA disagreement with Faddeev comes from the truncation
of the Sturmian expansion Eq.(\ref{sturm}). Eq.(\ref{wf3bad}) implies that all $\chi_i^{AD}(R)$ behave as the elastic $\chi_0^{AD}(R)$.
However, at higher beam energies, large excitation energies in the deuteron channel
are expected to affect the shape of the adiabatic distorted waves $\chi_i^{AD}(R)$.

\begin{table}[t!]
\caption{Percentage differences between the differential cross section at
the first peak of the distribution for the various formulations: $\Delta_{F-AD}$ comparing
Faddeev with Adiabatic, $\Delta_{F-Up}$ the effect of changing the energy at which the proton-target
interaction is calculated in the Faddeev, and finally $\Delta_{AD-rem}$ the effect of the remnant term in the adiabatic.
Also given is the beam energy $E$ (in MeV) and the angle $\theta$ at which the percentage difference was calculated (in degrees).}
\label{errors}
\begin{center}
\begin{tabular}{|ccc|cccc|}
\hline
reaction  & $E$ & $\theta$ & $\Delta_{F-AD}$ & $\Delta_{F-F2}$  & $\Delta_{AD-AD2}$  &$\Delta_{AD-rem}$ \\
\hline
$^{11}$Be(p,d)  &  5 & 1 	&  	-22.90	& 0.17 & -0.10	   		        & 1.36	 \\
		&  10& 1 	 &	-1.12		& 0.42	& 0.44		       & -1.13	 \\
		&  35& 1	 &	18.50		& 15.25	 & 10.9		       & 26.82	 \\
\hline
$^{12}$C(d,p )  &  7 & 20	 &	-9.08	& 1.61 & -0.42				& -3.74	 \\
		&  12 & 15 	&	-2.74		& 2.18	& -0.50			& -3.92	 \\
		&  56 & 1 	&	-14.26		& 13.95	 & 17.5			& 3.34	 \\
\hline
$^{48}$Ca(d,p)  &  19 & 8 	&	-6.12	& 7.16 & -0.64				& -0.74	 \\
		&  56 & 1 	&	-3.05		& 0.76	 & 46.4			& 0.57	 \\
		&  100& 0 	&	51.0		& 72.9	& -42.7		    & 9.73	 \\
\hline
\end{tabular}
\end{center}
\end{table}

\begin{figure}[!t]
{\centering \resizebox*{0.42\textwidth}{!}{\includegraphics{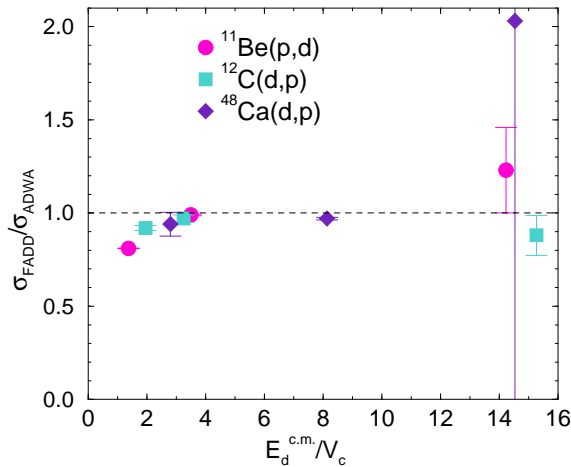}}} \\
\caption{\label{fig-ratio} (Color online) Ratio of Faddeev prediction for the cross section
at the first peak of the angular distribution versus the Adiabatic counterpart plotted in
term of  the deuteron energy in the c.m. over the Coulomb Barrier.}
\end{figure}

An attempt to systematize the results is done in Fig.\ref{fig-ratio} where the ratio of Faddeev cross sections
at the first peak of the angular distribution ($\theta$ from Table \ref{errors}) over the corresponding ADWA cross sections
is plotted as a function of $E_d^{cm}/V_c$, the c.m. energy of the deuteron ($E_d^{cm}$)
over a simple estimate of the Coulomb barrier $V_c=1.44 Z/(1.2 A^{1/3})$. We introduce
an error bar corresponding to the ambiguity in the choice of the energy at which the proton interaction
is calculated in the Faddeev (2nd column in Table \ref{errors}).  Figure \ref{fig-ratio} provides
a good illustration of the validity of ADWA for intermediate energies.
For the lowest energies, differences increase up to $20$\%  and they become particularly large for the highest
energies where the Hamiltonian ambiguities are also the largest.

In addition there should also be an error
due to the fact that in the Faddeev approach, the n-target interaction in the deuteron channel has no absorption for
the partial wave in which there exists a bound state. Based on previous experience, we expect this effect
to be smaller than the effect on the choice of $U_{pA}$ \cite{deltuva-E}.
Additional Faddeev calculations for the lighter targets,
introducing energy dependence in the n-target interaction in the partial wave corresponding to the
bound state, suggest the ambiguity on $U_{nA}$ is only comparable
to $\Delta_{F-F2}$  for $^{11}$Be(p,d) at 35 MeV.

We have also explored the dependence on the angular momentum $l$ of the neutron bound state as well as its binding energy $-S_n$.
For this purpose, we repeat the calculations for our best case $^{11}$Be(p,d)$^{10}$Be at 10 MeV. The neutron bound
state in $^{11}$Be has $l=0$ and $S_n=0.5$ MeV.
We first repeat the calculations changing the angular momentum to $l=1,2,3$ by adjusting the depth of the neutron bound state
interaction $V_{nA}$  to reproduce the binding $S_n=0.5$ MeV, while keeping all other interactions fixed. We find that as
$l$ increases, the difference between FADD and ADWA also increases: $\Delta(F-AD)=1$\% for $l=1$;
$\Delta(F-AD)=15$\% for $l=2$ and $\Delta(F-AD)=30$\% for $l=3$.

Next we repeat the calculations fixing $l=0$ and changing the depth of $V_{nA}$ to reproduce $S_n=2.0, 4.0$ and $6.0$ MeV.
Again, all other interactions are unchanged. The percentage difference relative to the Faddeev result is:
$\Delta(F-AD)=1$\% for $S_n=0.5$ MeV,  $\Delta(F-AD)=13$\% for $S_n=2.0$ MeV, $\Delta(F-AD)=18$\% for $S_n=4.0$ MeV
and $\Delta(F-AD)=23$\% for $S_n=6$ MeV.
In these calculations, remnant effects do not change significantly.
However, the deeper the neutron initial state, the more delocalized
is the $n-p$ system in the continuum, and one might expect the first term of the Sturmian expansion $\chi_0^{AD}(R)$ to
have the wrong radial behavior.
Whereas with increasing $l$ the ADWA diverges from the Faddeev in a rapid manner,
the differences with increasing binding appear to reach a saturation point around $\approx 20$ \%.
These results explain the findings in \cite{nunes11}.
In \cite{nunes11}, the discrepancy between ADWA and Faddeev was found to be much larger for $^{36}$Ar(p,d)
than for $^{34}$Ar(p,d) and $^{46}$Ar(p,d). The hole state created by pulling out a neutron
from $^{36}$Ar has both, large $l$ as well as large $S_n$.

\section{Summary and conclusions}

The finite range adiabatic model provides a practical method to analyze (d,p) and (p,d) reactions
that goes well beyond DWBA. One of the great advantages is that it uses nucleon optical potentials,
rather than deuteron optical potentials which are far more ambiguous. However it is by no means
the exact solution and in this work we quantify the errors introduced by comparing it to exact Faddeev AGS calculations. We choose three emblematic reactions $^{11}$Be(p,d)$^{10}$Be(g.s.),
$^{12}$C(d,p)$^{13}$C(g.s.) and $^{48}$Ca(d,p)$^{49}$Ca(g.s.) and span a large range of beam energies.
Overall, we find the agreement for reactions with beam energies around 10 MeV/u to be better than 10\%. These results bear
important implications for the reactions being measured at Isotope-Separator-on-Line facilities as well as for
the science program at the future Facility for Rare Isotope Beams.
The deviation of ADWA from Faddeev increases with the angular momentum of the neutron bound state
as well as with the separation energy of the neutron. The dependence on $l$ was found to be stronger
than that on the separation energy.

The comparison with Faddeev is limited in several ways. On one hand there are limitations with fixing
the three-body Hamiltonian. On the other hand there are technical issues that limit the number of cases
that can be studied. To address the first point, we have explored in detail the effect of different
choices for the proton target interaction, namely on the proton energy used to determine the optical potential.
The differences obtained are then used as a systematic uncertainty in the comparison. They are only important
at the higher energy, exactly where the performance of ADWA is at its worse. In this sense, the conclusion that ADWA fails
for energies larger than $E_d^{cm}/V_c \approx 12$ is not robust.

As mentioned in Section II,
there are several technical limitations in the present implementation of the Faddeev equations, namely
the AGS method in momentum space. In this respect we found $^{48}$Ca(d,p) at 19.3 MeV
 to be most challenging and
 close to the limit of present capabilities in obtaining converged results.

Since the finite range adiabatic model for (d,p)/(p,d) at intermediate energies
appears to be well suited to describe the few-body reaction dynamics for many cases of interest,
is easy to use and is not computationally intensive,
it is strongly desirable that the standard reaction codes incorporate this option in
a user-friendly manner.

In this work we did not consider target excitation, however it is understood that there will be many applications
in which target excitation is an integral part of the reaction mechanism. It would thus be worthwhile to
extend ADWA to include the inelastic channels in the formulation consistently.

\medskip

The work of F.M.N. was partially supported by the National Science Foundation
grant PHY-0555893, the Department of Energy through grant DE-FG52-08NA28552
and the TORUS collaboration DE-SC0004087.
The work of A.D. was partially supported by the
FCT grant PTDC/FIS/65736/2006.


\end{document}